\documentstyle[aps,pra]{revtex}

\def\la{{\langle}}
\def\ra{{\rangle}}

\def\tr{{\rm{Tr}}}
\newcommand{\beq}{\begin{equation}}
\newcommand{\eeq}{\end{equation}}
\newcommand{\beqa}{\begin{eqnarray}}
\newcommand{\eeqa}{\end{eqnarray}}
\newcommand{\da}{^\dagger}
\newcommand{\wh}{\widehat}

\newcommand{\intf}{\int_{-\infty}^\infty}
\newcommand{\into}{\int_0^\infty}

\draft
\begin{document}

\title{Time-of-arrival distributions from position-momentum and
energy-time joint measurements.}
\author{A. D. Baute$^1$, I. L. Egusquiza$^2$, J. G. Muga$^{1,3}$
and R. Sala-Mayato$^{1,4}$}
\address{$^1$ Departamento de F\'{\i}sica Fundamental
II, Universidad de La Laguna,
La Laguna, Tenerife, Spain}
\address{$^2$ Fisika Teorikoaren Saila,
Euskal Herriko Unibertsitatea,
644 P.K., 48080 Bilbao, Spain}
\address{$^3$ Departamento de Qu\'\i mica-F\'\i sica,
Universidad del Pa\'\i s
Vasco, Apdo. 644, 48080 Bilbao, Spain}
\address{$^4$ Institute for Microstructural Sciences, National Research
Council of Canada, Ottawa, Ontario K1A OR6, Canada}

\maketitle
\begin{abstract}
The position-momentum quasi-distribution obtained from an Arthurs and
Kelly joint measurement model is used to obtain indirectly an
``operational'' time-of-arrival (TOA) distribution following a
quantization procedure proposed by Kocha\'nski and W\'odkiewicz [Phys.
Rev. A {\bf 60}, 2689 (1999)].   This TOA distribution is not time
covariant.  The procedure is generalized by using other
phase-space quasi-distributions, and sufficient conditions are
provided for time covariance that limit the possible phase-space
quasi-distributions essentially to the Wigner function, which, however,
provides a non-positive TOA quasi-distribution. These problems are remedied
with a different quantization procedure which, on the other hand, does not guarantee
normalization. Finally an Arthurs and Kelly
measurement model for TOA and energy (valid also for arbitrary
conjugate variables when one of the variables is bounded from below)
is worked out.  The marginal TOA distribution so obtained, a distorted
version
of Kijowski's distribution, is time covariant, positive, and normalized.
\end{abstract}

\pacs{PACS: 03.65.-w\hfill EHU-FT/9913}

\section{Introduction}
In 1965 Arthurs and Kelly \cite{AK} proposed a model for the
simultaneous measurement of the position and momentum of a quantum
particle.  This model has been a valuable tool in many fundamental and
applied works, see e.g.
\cite{Leonh,Braun,Sten1,Wod2,App1,App2,Muy1,Muy2,Busch85}
and references
therein, and is present, albeit implicitly, in many others
\cite{AAU,OR,Wod,Hall}.
In the last few years, the advances in quantum optics have
led to experimental techniques and theoretical approaches devoted to
the simultaneous measurement of conjugate variables that have made
possible a practical realization of the original gedanken experiment
of Arthurs and Kelly \cite{Leonh,Braun,Sten1,Sten2,Agar}.
The measurement process is modeled
by a Hamiltonian that includes degrees of freedom of the particle and
two pointers. It is constructed by adding two sudden interaction terms that,
when acting separately,
provide  impulsive (or von Neumann) measurements \cite{VN}
of the position and the momentum of the particle at time $t$,
\beq\label{Hami}
\wh{H}_{AK}=\delta(t)(\wh{\pi}_P \wh{p}+\wh{\pi}_X \wh{x})\,.
\eeq
All other terms are neglected, in particular the ones corresponding to the free
dynamics of pointers and particle.  In each of these two
partial interaction Hamiltonians the operator for the particle
property to be measured (position $\wh{x}$, resp.  momentum $\wh{p}$)
is multiplied by the conjugate operator, $\wh{\pi}_X$, resp.
$\wh{\pi}_P$, of the associated pointer observable $\wh{\mu}_X$, resp.
$\wh{\mu}_P$, so that for each of the ordinary von Neumann
measurements the state of the pointer variable is displaced
proportionally to each eigenvalue \cite{VN}.
However, the combination of the two interaction terms in a single
Hamiltonian implies a mutual disturbance of the two measurements.

In the combined (joint) measurement the displacement of the two
commuting pointer variables $\mu_X$ and $\mu_P$ can be described by a
true joint distribution, $\rho(\mu_X,\mu_P)$.  It is well known that a
unique position-momentum distribution for $x$ and $p$ (positive,
bilinear in the wave function, and with the correct marginals) cannot
be defined in quantum mechanics, but there are many possible
quasi-distributions.  The Arthurs and Kelly model provides a simple
operational realization of a family of joint
position-momentum quasi-distributions in terms of the commuting
pointer positions as \cite{Sten1}
\beq\label{frho}
F(x,p)\equiv\rho(\mu_X=x, \mu_P=p)\,.
\eeq
In fact Arthurs and Kelly considered one particular set of states for the
apparatus
that makes $F$ a ``Husimi function'' \cite{Sten1}; however,
more general apparatus
states are possible.  The resulting family of quasi-distributions
has been discussed by several authors \cite{Davies,Wod}.

The Arthurs and Kelly process also provides a natural way of
quantizing classical functions of position and momentum.  In
particular, for the free motion case, Kocha\'nski and W\'odkiewicz have
recently defined ``operationally'' a time-of-arrival  (TOA) distribution
{}from the phase-space distribution $\rho$ \cite{KW99}.
As in (\ref{frho}), the idea is to
use the commuting variables $\mu_X$ and $\mu_P$, instead of the
non-commuting particle position and momentum, in the classical
expression that defines the arrival time.

The theoretical treatment of time observables is an important loose
end of the standard quantum mechanical formalism.  Among these
observables, the time of arrival has been investigated  using
many different approaches, as reviewed in \cite{Mugarev} - for more
recent works see
\cite{MLP,Finkelstein,Toller,Kijo99,ORU,Delgado,MPL,EM,BEMS,LJPU,Galapon}.
They may be classified
according to their intrinsic (ideal) or operational nature.  Intrinsic
models consider the particle on its own, without
any external influence other than the potential in which it moves, and
provide ideal quantities that do not depend on any other degree of
freedom.  An example is the distribution of Kijowski \cite{Kijowski},
which satisfies in an unique manner a number of properties motivated
by its classical analog, or the distribution obtained within the
causal theory of Bohm \cite{Bohm}.  Operational models, instead,
include extra degrees of freedom for the measuring device explicitly
in the Hamiltonian \cite{KW99,AOPRU,Halliwell}, or implicitly by means of a
non-unitary evolution law (e.g.,
with effective complex Hamiltonians
\cite{MBM,Mugarev}). Notice that the word ``operational'' has a rather
different meaning in the usage of some authors \cite{Busch}, but we shall use
it in the sense described.
This paper investigates the properties (in particular covariance) of
operational time-of-arrival distributions obtained by means of Arthurs
and Kelly measurements, both of position and momentum and of the
conjugate variables energy and time of arrival.  In order to attain a
broader perspective, we also consider and put forward other possible
TOA distributions, based on different phase-space quasi-distributions,
not directly derived from an Arthurs and Kelly measurement,
and look into the relation between operational and ideal quantities.

In Sec. II we shall for completeness rederive the phase-space
distribution $\rho(\mu_X,\mu_P)$,
using the treatment of the Arthurs and Kelly
model given by Appleby \cite{App1,App2}.  In Sec. III we show that the
TOA  distribution thus
obtained, following Kocha\'nski and W\'odkiewicz, is not covariant.
We analyze in the following section  alternative distributions, with
particular
emphasis on the property of covariance, and conclude that, within a
broad family of phase space functions \cite{Cohen1,Cohen2}, only the
Wigner function gives a covariant TOA distribution when following the
recipe proposed in \cite{KW99}.  This covariant distribution, however,
is not positive, and this defect, in turn, leads us to investigate
other quantization recipes
to obtain covariant and positive distributions, a task in which we succeed,
using as before the simultaneous Arthurs and Kelly
measurement of
position and momentum. The result however is not normalizable.
Yet
another option in order to obtain covariant TOA distributions is
to perform an Arthurs and Kelly measurement for the conjugate
variables time of arrival and energy, as we show in Sec. V. This is
a simple model for the important experimental need to know the energy
{\it and} the time when the particles arrive at detectors.  The
results follow closely but not exactly the standard Arthurs and Kelly
process for position and momentum, and we are again successful in
obtaining covariant TOA distributions, now correctly normalized.
In the final section of
conclusions we indicate other possible extensions of our results.

\section{The Arthurs and Kelly measurement for position
and momentum}
Recently, Appleby has studied thoroughly the concepts of accuracy and
disturbance in the Arthurs and Kelly model \cite{App1,App2}.  We shall
show how to recover, following his analysis, the results for the
probability $\rho(\mu_X,\mu_P)$, previously obtained in a concise
manner by Kocha\'nski and W\'odkiewicz.  The interest of this detailed
rederivation is that the meaning of the apparatus dependent ``window''
function is made explicit.  The window function provides $\rho$, when
convolved with the particle's state, see (\ref{rhoxp}) below.  Our
derivation also applies to a class of apparatus states more general
than that in \cite{KW99}.

The operators for the six variables involved in the  Arthurs and Kelly
process satisfy the commutation relations
\beq\label{ihxp}
[\wh x,\wh p]=[\wh {\mu}_X,\wh {\pi}_X]=[\wh {\mu}_P,\wh {\pi}_P]=
i\hbar\,.
\eeq
Any other pair of these operators commutes (i.e.
$\wh{\mu}_P$ and $\wh{\mu}_X$, $\wh{x}$ and $\wh{\pi}_X$, etc...).
The unitary evolution operator describing the measurement process is
\beq\label{U}
\wh{U}_{XP}=e^{-\frac{i}{\hbar}(\wh{\pi}_P \wh{p}+\wh{\pi}_X \wh{x})}\,.
\eeq
There is no explicit reference to the measurement instant $t$ here.
$\wh{U}_{XP}$  may be regarded as the evolution
operator connecting the states
of the system  before and after a sudden interaction
(\ref{Hami}). It is also possible
to interpret it as the evolution operator that gives the
final state for an interaction $K(\wh{\pi}_P \wh{p}+\wh{\pi}_X \wh{x})$
acting during a time $\Delta t=1/K$, with $K$ sufficiently large so that
all other terms in the Hamiltonian can be
neglected during the measurement time $\Delta t$ \cite{AK}.

If the initial state of the global particle+apparatus system is
given by the product
state $|\psi\otimes\psi_{ap}\ra$, the probability distribution for the
result of the measurement takes the form
\beq\label{key}
\rho(\mu_X, \mu_P)=\intf dx\, |\la x, {\mu}_X, {\mu}_P|\wh{U}_{XP}|
\psi\otimes\psi_{ap}\ra|^2\,.
\eeq
This is the key expression that relates the probability distribution
of the pointer variables $\mu_X$ and $\mu_P$ to the particle's initial
state.  Let us emphasize that $\rho(\mu_X,\mu_P)$ is a {\it true}
probability distribution of its commuting variables, $\mu_X$ and
$\mu_P$; however, when interpreted as a function of the {\em quantum
mechanical} particle
variables, namely as $F(x,p)\equiv \rho(\mu_X=x, \mu_P=p)$, it is to be
understood more properly as a quasi-distribution. This comes about
because the marginal distributions of $F$ for $x$ and $p$ are not
the quantum mechanical ones for the state of the particle,
even though they are correctly normalized and positive.

To arrive at an explicit form for such a distribution and to describe
the experimental errors in the measurement
process, Appleby introduces the initial and final ``Heisenberg
picture'' operators $\wh{\it O}_i=\wh{\it O}$ and $\wh{\it O}_f={\wh
{U}_{XP}\da}\wh{\it O}\wh {U}_{XP}$, where $\wh O$ can be any of the
six operators in (\ref{ihxp}).  In terms of these operators several
``errors'' are defined.  In particular the {\it
retrodictive error} operators,
\beqa\label{ret}
\wh{\epsilon}_{Xi}&=&\wh{\mu}_{Xf}-\wh{x}_i\,,
\nonumber\\
\wh{\epsilon}_{Pi}&=&\wh{\mu}_{Pf}-\wh{p}_i\,,
\eeqa
provide the accuracy with which the result of the measurement reflects
the state of the system {\it before} the measurement was carried out;
and the {\it predictive error} operators,
\beqa\label{pre}
\wh{\epsilon}_{Xf}&=&\wh{\mu}_{Xf}-\wh{x}_f\,,
\nonumber\\
\wh{\epsilon}_{Pf}&=&\wh{\mu}_{Pf}-\wh{p}_f\,,
\eeqa
give the accuracy with which the result of the measurement reflects
the state of the system {\it after} the measurement.

Expressions for the final operators in (\ref{ret}) and (\ref{pre}),
\beqa\label{fin}
\wh{x}_f &=& {\wh {U}_{XP}\da}\wh{x}\,\wh {U}_{XP} = \wh{x}+\wh{\pi}_P\,
,
\nonumber\\
\wh{p}_f &=& {\wh {U}_{XP}\da}\wh{p}\,\wh {U}_{XP} = \wh{p}-\wh{\pi}_X\,
,
\nonumber\\
\wh{\mu}_{Xf} &=& {\wh {U}_{XP}\da}\wh{\mu}_X\wh {U}_{XP} = \wh{\mu}_X+
\wh{x}+
\frac{1}{2}\wh{\pi}_P\,,
\nonumber\\
\wh{\mu}_{Pf} &=& {\wh {U}_{XP}\da}\wh{\mu}_P\wh {U}_{XP} = \wh{\mu}_P+
\wh{p}-
\frac{1}{2}\wh{\pi}_X\,,
\eeqa
are easily obtained using the commutation relations (\ref{ihxp}) and the
following relation,  valid
for two arbitrary operators $\wh A$ and
$\wh B$,
\beq\label{rela}
e^{\gamma \wh A}\wh {B}e^{-\gamma \wh A}=\wh{B}+\gamma [\wh{A},\wh{B}]+
\frac{\gamma^2}{2!}\left[\wh{A},[\wh{A},\wh{B}]\right]+
\frac{\gamma^3}{3!}\left[\wh{A},\left[\wh{A},[\wh{A},\wh{B}]\right]\right]+
....
\eeq
The distribution of measured values (\ref{key}) can finally be written after
some algebra as
a convolution in phase space \cite{App2}
\beq\label{rhoxp}
\rho(\mu_{X},\mu_{P})=\intf\!\intf dx\,dp\,\,
W_{\epsilon i}(\mu_{X}-x, \mu_{P}-p)
w(x,p)\,,
\eeq
where $w(x,p)$ is the Wigner function of the initial state of the
particle (just before the measurement takes place),
\beq\label{wsyi}
w(x,p)=\frac{1}{h}\intf dy\, e^{\frac{i}{\hbar}py}
\bigg< x-\frac{y}{2}\bigg|\psi\bigg>\bigg<\psi\bigg|x+\frac{y}{2}\bigg >\,,
\eeq
and the apparatus dependent ``window'', or ``filter'' function in phase
space,
\beq\label{wei}
W_{\epsilon i}(\mu_{X}-x, \mu_{P}-p)=\frac{1}{h}\intf dy\,
e^{\frac{-i}{\hbar}\epsilon_{Pi}y}
\bigg < \epsilon_{Xi}-\frac{y}{2}\bigg|\wh{\varrho}_{\epsilon_i}\bigg|
\epsilon_{Xi}+\frac{y}{2}\bigg >\,,
\eeq
where $\epsilon_{Pi}=\mu_P-p$, and $\epsilon_{Xi}=\mu_X-x$,
is the Wigner function corresponding to $\wh{\varrho}_{\epsilon_i}$,
the reduced initial apparatus state
density operator for retrodictive errors,
\beq\label{red}
\la \epsilon_{Xi}|\wh{\varrho}_{\epsilon_i}|\epsilon'_{Xi}\ra =
\intf d\epsilon_{Xf}\,\la \epsilon_{Xi},\epsilon_{Xf}|\psi_{ap}\ra
\la \psi_{ap}|\epsilon'_{Xi},\epsilon_{Xf}\ra\,.
\eeq
The trace is taken over the predictive error of position.  Note that
the apparatus state has been represented in the basis of the complete
set of commuting operators (for the apparatus space) of retrodictive
and predictive position errors,
$\wh{\epsilon}_{Xi}=\wh{\mu}_X+\wh{\pi}_P/2$, and
$\wh{\epsilon}_{Xf}=\wh{\mu}_X-\wh{\pi}_P/2$ (that refere exclusively
to the apparatus, and whose conjugate momenta
are $-\wh{\epsilon}_{Pi}$ and $\wh{\epsilon}_{Pf}$ respectively).
By changing the distribution of retrodictive errors it is
possible to obtain a family of operational phase space distributions
\cite{Leonh,Dav,Busch}.

\section{Time-of-arrival distribution: {\it indirect} approach}

By assuming a separable, pure state form for
$\wh\varrho_{\epsilon_i}=|\phi\ra\la\phi|$ in (\ref{wei}),
and using expression (\ref{wsyi}) for
$w(x,p)$ in
Eq. (\ref{rhoxp}), the distribution of measured values
of position and momentum can be written as
\beq\label{spectro}
\rho(\mu_{X},\mu_{P})=\frac{1}{h}\left|\intf dx'\, \phi^*(\mu_{X}-x')\psi(x')
e^{-i\mu_{P}x'/\hbar}\right|^2\,,
\eeq
where the ``filter'' or ``window'' function $\phi(\mu_{X}-x')$ is the
retrodictive error wave function of the apparatus.  (It is to be noted
that while the form of (\ref{spectro}) depends on the assumed
factorized structure of the reduced retrodictive error density
operator, (\ref{rhoxp}) is more general and does not require such
structure.)  Here we may see that the quasi-distribution functions
$F(x,p)=\rho(\mu_X=x, \mu_P=p)$ are, in the language of time-frequency
analysis \cite{Cohen2}, nothing but ``spectrograms'', that is, the
square modulus of Fourier transforms of the particle state multiplied
by an apparatus-dependent window function selecting a limited spatial
region.  Clearly, the transform reflects the combined properties of
the particle state and the window function.  A natural condition to
impose on $\phi(\epsilon_{Xi})$ is that it be centered at $0$, so that
the average retrodictive error of position is zero, and the spatial
regions where $\psi$ and $\rho$ have significant values are in good
agreement, at least on average (since otherwise the pointer would be
displaced with respect to the particle's position).  Similarly, for
the momentum representation version of $(\ref{spectro})$ we shall
demand that the Fourier transform of the window function be also
centered at zero retrodictive error of momentum.  In summary, and
using the language of \cite{App1}, we shall assume that the measurement
is ``retrodictively unbiased''.

Note that in the argument of $\phi$ the sign of $x$ is different from
the one in \cite{KW99}.  This can be traced back to different sign
conventions for the momentum pointer and the commutation relation with
its conjugate variable.  Both conventions lead to equivalent results
if appropriate sign changes are taken into account.  With the present
convention, and for unbiased measurements, the pointers are located
(on average) at the particle's average position and momentum, and thus
$\rho(\mu_X,\mu_P)$ tracks directly (of course with the unavoidable
distortions inherent to the joint measurement) the particle's position
and momentum distributions.  In \cite{KW99}, instead, the position
pointer and the particle's position have opposite signs when the
filter function is centered at $0$, see Eq.  (21a) of \cite{KW99} with
$q_0=0$.

Kocha\'nski and W\'odkiewicz have used the operational phase space
distribution to define, {\it indirectly}, a TOA
distribution for the free particle \cite{KW99}. The basic idea is to work
with the commuting
operators $\wh{\mu}_X$ and $\wh{\mu}_P$ instead of
$\wh{x}$ and $\wh{p}$.
A classical particle with position $x$ and momentum $p$ at time $t$ takes a time 
$T=-mx/p$, measured from $t$, to arrive at the origin. This
motivates the definition of a distribution of arrival times by means
of the following average in the ``observed''  phase space $\mu_X,\mu_P$,
\beqa\label{Prw}
\Pi_{KW}(T;t)&\equiv&\bigg<\delta \left(\frac{m\mu_X}{\mu_P}+T\right)\bigg>
=\intf\!\intf d\mu_X\,d\mu_P\, \delta
\left(\frac{m\mu_X}{\mu_P}+T\right) \rho (\mu_X,\mu_P,t)
\,,
\eeqa
where $- m \mu_X /\mu_P$ is used instead of the classical
expression.
(Note that $T$ is the time interval from $t$ to the arrival instant $t+T$.)

As an example, and in order to have a better grasp of the properties
of this distribution, assume, as in \cite{KW99},
that the particle initial state is a
Gaussian wave function with initial mean position $x_0$, initial mean
momentum $\hbar k_0$ and position ``width'' $\delta$ (square root of the
variance),
\beq
\psi(x,t)=\left(\frac{2\delta^2}{\pi}\right)^{1/4}
\frac{e^{-\frac{1}{4}k_0^2 \delta^2}}{\sqrt{\delta^2 +\frac{2it\hbar}{m}}}
e^{\left[\frac{1}{2}k_0\delta^2 +i(x-x_0)\right]^2/(\delta^2 +\frac{2it\hbar}
{m})}\,,
\eeq
and that the initial ``apparatus filter-function''
is given by an unbiased  Gaussian
with width $\sigma$,
\beq
\phi(\mu_X-x)=\left(\frac{2}{\pi\sigma^2}\right)^{1/4}
e^{-(\mu_X-x)^2/\sigma^2}\,.
\eeq
For the example at hand the distribution $\rho (\mu_{X},\mu_{P},t)$
can be written explicitly:
\beqa\label{model}
\rho (\mu_X,\mu_P,t)&=&\frac{\sigma\delta}{2\pi\hbar}
\frac{1}{\sqrt{\frac{4t^2\hbar^2}{m^2} +(\delta^2 +\sigma^2)^2}}
\exp\left(-\frac{1}{2\hbar^2}\frac{\delta^2\sigma^2(\delta^2
+\sigma^2)(\mu_P-\hbar k_0)^2} {\frac{4t^2\hbar^2}{m^2} +
(\delta^2 +\sigma^2)^2}\right)\nonumber\\
&\times&\exp\left(\frac{-2[\delta^2(x_0+\frac{\hbar k_0t}{m}-\mu_X)^2
+\sigma^2(x_0+\frac{\mu_Pt}{m}-\mu_X)^2]} {\frac{4t^2\hbar^2}{m^2} +
(\delta^2 +\sigma^2)^2}\right)\,.
\eeqa
Inserting this expression into (\ref{Prw}) we obtain
\beqa\label{PKW}
\Pi_{KW}(T;t)&=&\frac{\hbar}{2\pi m}
\frac{\delta\sigma\sqrt{\frac{4t^2\hbar^2}{m^2} +(\delta^2
+\sigma^2)^2}}{\Delta(T,t)}
\exp\left(\frac{-2k_0^2\left[\delta^2T_{cl}^{2}(t)
+\sigma^2T_{cl}^{2}(0)+\Delta(0,0)\right]}
{\frac{4t^2\hbar^2}{m^2} +(\delta^2 +\sigma^2)^2}\right)\nonumber\\
&\times&\left(1+\sqrt{\pi}\xi(T,t) e^{\xi^2(T,t)}\Phi[\xi(T,t)]\right)\,.
\eeqa
where $\Phi$ is the error function and the following
symbols have been used:
\beqa
\Delta(T,t) & =& \frac{\sigma^2(T+t)^2\hbar^2}{m^2}
+\frac{\delta^2T^2\hbar^2}{m^2} +
\frac{1}{4}\delta^2\sigma^2(\delta^2+\sigma^2)
\\
T_{cl}(t) & = & -\frac{x_0+\frac{k_0t\hbar}{m}}{k_0}\,,
\\
\xi(T,t) & = & \frac{2k_0[\Delta(0,0)+ \hbar (
\sigma^2(T+t) T_{cl}(0)
+\delta^2 T T_{cl}(t))/m]}
{\sqrt{2\Delta(T,t)}\sqrt{(\delta^2+\sigma^2)^2+4t^2\hbar^2/m^2}}\,.
\eeqa
This distribution is shown in Figure $1$, where its lack of covariance is
evident. According to the expression (\ref{PKW}), the
probability for
arriving at instant $T+t$, when an interval  $T$ has passed after the reference 
time $t$, is not equal  to the probability for arriving 
at the same instant ($T+t$), when an interval $T-t'$ has passed after the 
reference time $t+t'$, 
\beq\label{cov}
\Pi_{KW}(T;t)\neq \Pi_{KW}(T-t';t+t')\,.
\eeq
Covariance is however a basic physical requirement for any good
quantum time-of-arrival distribution
\cite{Werner,Busch,MLP}.  It simply means that the number of arrivals
predicted for a particular fixed instant ($T+t$ in \ref{cov})  should be 
a constant quantity independent of $t'$, i.e., on the reference time used in making the 
prediction. A good
apparatus should  give a stable, fixed answer, independent
of the instant that we have switched it on. Lack of covariance implies
that different predictions are given about the number of arrivals at the
{\it same} instant of time depending on the reference time chosen
to make the question (that corresponds here to the
Arthurs and Kelly measurement).
We may wonder whether a time distribution obtained from other phase-space
representations of the particle state can be covariant with respect to time
translations.  This study is carried out in the next section.

\section{Covariance in time}
The question we shall address first is whether any of the many possible
representations of the quantum state, other than the spectrogram, does
provide a covariant time-of-arrival distribution following the recipe
shown in (\ref{Prw}). The process of Arthurs and Kelly
associates states of a particle with probability
distributions on the phase space of the particle, via Eq. (\ref{frho}).
This association, in
the case of Arthurs and Kelly measurements, is done through the filter
function, derived from the state of the measuring apparatus. There are
however many other ways of building up such a pairing between
particle state and particle phase space. (Only some of
them have a simple  operational interpretation in terms of a measurement
model).

A very broad class of quantum quasi-probability distributions $F$ of
position and momentum was studied and defined by Cohen
\cite{Cohen1,Cohen2}, the one in
Eq. (\ref{spectro}) being a particular case (see below).
In Cohen's approach each of the distributions $F$ is obtained
with a different kernel $\chi$
{}from the  density operator $\wh\varrho$ of the particle,
see Eq.(\ref{Frho}) in the appendix;
and for  each of these kernels a quantization rule is defined,
Eq. (\ref{Gg}),
such that the  expectation values
can be equally obtained by means of phase space integrals or operator
traces, Eq. (\ref{expec}). In particular,
\beq\label{Pif}
\Pi_\delta(T;t;[\chi])\equiv\bigg<\delta\left(\frac{mx}{p}+T\right)\bigg>
=\intf\!\intf dx\,dp\,\, \delta\left(\frac{mx}{p}+T\right) F(x,p,t;[\chi])
=
\rm{Tr}[\wh{\varrho}(t)\,\wh{\delta}_\chi(T)]\,,
\eeq
where $\wh{\delta}_\chi(T)$ is a shorthand notation for the operator
corresponding
to the classical ``function'' $\delta(mx/p+T)$ by means of the
``$\chi$-quantization rule'',
\beq
\wh{\delta}_\chi(T)\equiv\frac{1}{4\pi^2}\intf\!\intf\!\intf\!\intf
dx\,dp\,d\theta\,d\tau\,
\delta(T+xm/p)\chi(\theta,\tau)\exp[-i(\theta(x-\wh{x})+\tau(p-\wh{p})]\,.
\eeq
Covariance in time of the arrival time distribution means
that $\Pi_\delta(T;t;[\chi])$ should be equal to
\beqa
\Pi_\delta(T-t';t+t';[\chi])&=&\tr[\wh{\varrho}(t+t')\wh{\delta}_\chi(T-t'
)]
\\
&=&
\tr\left[\wh{\varrho}(t)
e^{i\wh{H}t'/\hbar}\wh{\delta}_\chi(T-t')e^{-i\wh{H}t'/\hbar}\right]
\eeqa
for all $\wh\varrho(t)$. Thus the covariance condition
can be expressed in operator form,
\beq\label{covaop}
\wh{\delta}_\chi(T)
=
e^{i\wh{H}t'/\hbar}\wh{\delta}_\chi(T-t')e^{-i\wh{H}t'/\hbar}\,.
\eeq
Our next task is to find kernels $\chi$ that fulfill (\ref{covaop}).
To this end we shall work out the momentum representation of the
two sides of (\ref{covaop}) for an arbitrary $\chi$. For the left hand
side we find:
\beqa\label{del1}
\la p'|\wh{\delta}_\chi(T)|p''\ra&=&
\frac{1}{2\pi h m}\intf\!\intf dp\,d\tau |p|\,\chi\left[\frac{p'-p''}{\hbar},
\tau\right]
e^{-ip(\tau-T(p'-p'')/m\hbar)}e^{i\tau(p'+p'')/2}
\\
&=&
\frac{e^{i({p'}^2-{p''}^2)T/(2m\hbar)}}{2\pi h m}
\intf\!\intf dp\, d\sigma\, |p|\,\,
\chi\left[\frac{p'-p''}{\hbar}, \sigma+\frac{T(p'-p'')}{m\hbar}\right]
e^{i\sigma[p+(p'+p'')/2]}\,.
\eeqa
Use has been made of the relation
\beq
e^{i(\theta\wh{q}+\tau\wh{p})}=e^{i\hbar\theta\tau/2}
e^{i\theta\wh{q}}e^{i\tau\wh{p}}\,,
\eeq
and of the change of variable $\sigma=\tau-T(p'-p'')/m\hbar$.
Operating similarly, the momentum representation of the
right hand side of (\ref{covaop}) takes the form
\beqa
\la p'|e^{i \wh{H} t'/\hbar}\wh{\delta}_\chi(T-t')e^{-i \wh{H}t'/
\hbar}|p''\ra
&=&\frac{e^{i({p'}^2-{p''}^2)T/(2m\hbar)}}{2\pi h m}
\nonumber\\
\label{del2}
&\times&\intf\!\intf dp\, d\sigma\,
|p|\,\, \chi\left[\frac{p'-p''}{\hbar},
\sigma+\frac{(T-t')(p'-p'')}{m\hbar}\right]
e^{i\sigma[p+(p'+p'')/2]}\,.
\eeqa
If (\ref{del2}) has to be equal to (\ref{del1}) for all $t'$, the kernel
function $\chi(\theta, \tau)$ must be independent of its second argument.
Since the only set of kernels worth considering are those that preserve the
normalization of the state, i.e., such that $\chi(0,0)=1$
\cite{Cohen1,Cohen2}, independence of $\tau$ implies $\chi(0,\tau)=1$,
which is a sufficient condition to provide a correct marginal
distribution for $p$. Therefore,  independence on $\tau$ limits the set of
possible $\chi$ kernels rather strongly, in particular the spectrogram
does not generally belong to this class, since it does not
generally satisfy the marginals.
Its  kernel has the following form, see (\ref{spectro}),
\beq
\chi(\theta,\tau)=\intf dy\,\phi^*(y-\tau\hbar/2)\phi(y+\tau\hbar/2)
e^{i\theta y}\,.
\eeq
This may become independent of $\tau$ in the
limit of a very
flat $\phi$ function, namely, for a
vanishing retrodictive error of momentum. This is not a
desirable limit for our purposes:
the position becomes so
imprecise, that $\Pi_{KW}(T)$ also tends to vanish in that limit, as can be
seen, for example, by taking $\sigma\to\infty$ in Eq. (\ref{PKW}).

Restricting ourselves to the $\chi$ functions
that provide an $F$ with the two correct marginals (this requires
$\chi(\theta,0)=\chi(0,\tau)=1$ \cite{Cohen1,Cohen2}, and excludes
the spectrograms), the independence on
$\tau$ can only be satisfied by $\chi=1$, which is the kernel that
corresponds to the Wigner function and the associated Weyl
quantization rule.  Furthermore, in the set of the kernels which provide
scale-invariance \cite{Cohen2}, characterized by being functions of
the product $\tau\theta$, $\chi=1$ is again the only possible case that
presents covariance.

The covariant TOA distribution obtained with the Wigner function
and (\ref{Pif}) takes the
form
\beqa\label{covq}
\Pi_{\delta,Wigner}(T;t)&=&
\frac{1}{hm}\intf\!\intf dp'dp''\bigg|\frac{p'+p''}{2}\bigg|
\la p'|\wh\varrho(t+T)|p''\ra\nonumber\\
&=&\Pi_{\delta,Wigner}(T -t';t+t')\,.
\eeqa
Let us compare this result with the flux at the origin,
$J=\intf dp\,
w(0,p,t)\frac{p}{m}$.  By substituting (\ref{wsyi}), and performing
several integrals, this takes the form
\beq
J=\frac{1}{hm}\intf\!\intf dp'dp''\left(\frac{p'+p''}{2}\right)
\la p'|\wh\varrho(t)|p''\ra\,.
\eeq
The only difference with (\ref{covq}) is the presence
or absence of the absolute value in the half-sum, the two results
being equal for states without negative momentum components.  However,
a quantum mechanical state composed by positive momenta is compatible
with a negative value of $J$ (backflow) at certain times and positions
\cite{Allc,Brac,MPL}.  Therefore, the ``time-of-arrival
quasi-distribution'' obtained by means of Wigner's function satisfies
the covariance under time translations, but fails to be a positive
distribution, which is also an important requirement for a good
time-of-arrival distribution.

\subsection{Kijowski's distribution}
It is also of interest to investigate whether there is a kernel
function $\chi$ such that
the corresponding distribution $\Pi(T)$ obtained via  (\ref{Pif})
is equal to the covariant distribution of Kijowski,
\beq\label{Kijo}
\Pi_K(T;t)=\tr\left[\wh{\varrho}(t)
\left(\sum_\alpha|T,\alpha\ra\la T,\alpha|
\right)\right]\,.
\eeq
Here $|T,\alpha\ra$ are the (generalized) eigenfunctions of the
Aharonov-Bohm
time-of-arrival operator (see below),
\beq\label{p|T}
\la p|T,\alpha \ra=\left( \frac{|p|}{mh}\right)^{1/2} e^{ip^{2}T/2m\hbar}
\Theta(\alpha p)\,,
\eeq
and $\alpha=\pm$ is the degeneracy index associated with positive or
negative momentum.

We look for a kernel $\chi$ such that $\wh{\delta}_\chi(T)=\sum_\alpha
|T,\alpha\ra\la T,\alpha|$. Working again in momentum representation,
\beq\label{condi}
\la p'|\wh{\delta}_\chi(T)|p''\ra=\Theta(p'p'')\frac{(p'p'')^{1/2}}{mh}
e^{i({p'}^2-{p''}^2)T/(2m\hbar)}e^{i(p'-p'')X/\hbar}\,.
\eeq
Comparing with (\ref{del1}) we see that $\chi$ need not depend on its second
argument, as should have been expected since $\Pi_K$ is covariant.
The integrals in (\ref{del1}) can be then carried out, and the requirement
(\ref{condi}) becomes
\beq
\chi\left(\frac{p'-p''}{\hbar}\right)
=\Theta(p'p'')(p'p'')^{1/2}\left|\frac{2}{p'+p''}\right|\,.
\eeq
Changing variables to $\nu=p'-p''$ and $\eta=(p'+p'')/2$, it is clear
that no  function of $\nu$ can satisfy this equation,
since the right hand side also depends on $\eta$.  It follows that
Kijowski's distribution cannot be obtained by such an extension of the
procedure of Kocha\'nski and W\'odkiewicz.

\subsection{Alternative quantization of time of arrival}

Up to now we have used position-momentum representations of the particle
state to define time distributions by means of (\ref{Pif}), i.e., by
an extension of the procedure proposed by Kocha\'nski and W\'odkiewicz.
Let us now show that this does not exhaust all possible
quantizations of the classical distribution, and in fact that it is possible
to do much better with respect to covariance and positivity.
Consider the following quantities,
\beq\label{PiJ}
\Pi_{\tilde{J}}(T;t;[\chi])\equiv
\intf dx\intf dp \left|\frac{p}{m}\right|\delta(x)\,
F(x,p;T+t;[\chi])\,,
\eeq
where the subscript $\tilde{J}$ indicates that in the classical
version of this expression, where $F$ would be a true phase space joint
distribution, $\Pi_{\tilde{J}}$ represents the positive minus the
negative fluxes (the positive flux is defined by
$\int_0^\infty dp\,F(0,p)p/m$, and the negative flux by
$\int_{-\infty}^0dp F(0,p)p/m$).
This is apparently quite different from (\ref{Pif})
but, in fact, in the context of classical mechanics of the free
particle, the two
expressions are equivalent, as can be seen by using Liouville's
theorem and the trajectory equation for free motion.  However, their
quantizations are not the same in general.
 Note in particular that in
(\ref{PiJ}) the time dependence has been put entirely in the state and
not in the observable.  An important consequence is that the
functions $\Pi_{\tilde{J}}(T;t;[\chi])$ are time covariant {\em for all}
$\chi$.
A case where (\ref{Pif})
and (\ref{PiJ}) {\it are} equal is $\chi=1$, because the propagator
of the Wigner function for free motion is just the classical
propagator. In general,  (\ref{PiJ}) can be understood as a
``covariantization'' of (\ref{Pif}): they coincide for $T=0$, and
(\ref{PiJ}) is covariant by construction, which means that
\beq
\Pi_{\tilde{J}}(T;t;[\chi])=\Pi_{\delta}(0;t+T;[\chi])
\eeq
for all $t$ and $T$, and function $\chi$.

The difference  between (\ref{PiJ}) and (\ref{Pif}) is actually more
profound: whereas in (\ref{Pif}) we are performing {\it quantization of a
family} of classical functions, parameterised by $T$, namely,
$\delta(mx/p+T)$, (\ref{PiJ}) pertains to a {\it family of 
quantizations} for the {\it same} classical function,
$\left|\frac{p}{m}\right|\delta(x)$, each of the quantizations being
parameterised by $T$.
Since the value of Kijowski's distribution for a given $T$ is actually
the trace with the state of a suitable
operator, it seems unlikely that it can be reproduced within a scheme
so radically different in conceptual terms as a family of
quantizations, and, in fact, using a procedure similar to the one in
the previous subsection, one can prove that Kijowski's distribution
cannot be obtained from the quantization of
$\left|\frac{p}{m}\right|\delta(q)$ with any $\chi$ within the class
being considered \cite{class}.

Even so, positive TOA distributions can be obtained
{}from expression (\ref{PiJ}) for suitable quantization functions
$\chi$.  Indeed, a sufficient condition is that $F[\chi]$ be positive.
In particular, the spectrogram is positive and provides a covariant
TOA distribution via (\ref{PiJ}).  For the example worked out before,
see (\ref{model}),
\beqa
\Pi_{\tilde{J}}(T;t)
&=&\frac{\hbar}{2\pi m}\frac{\delta\sigma\sqrt{\frac{4(t+T)^2\hbar^2}{m^2}
+(\delta^2
+\sigma^2)^2}}{\Delta(0,t+T)}
\exp\left(\frac{-2k_0^2[\delta^2 T_{cl}^{2}(t+T)
+\sigma^2T_{cl}^{2}(0)+\Delta(0,0)]}
{\frac{4(t+T)^2\hbar^2}{m^2} +(\delta^2 +\sigma^2)^2}\right)\nonumber\\
&\times&\left(1+\sqrt{\pi}\xi(0;T+t) e^{\xi^2(0;T+t)}\Phi[\xi(0;T+t)]
\right)\,.
\eeqa
This distribution is represented in Figure 1.
The problem is that, due  to the asymptotic dependence $\sim 1/|T|$
for large $|T|$,
it  cannot be normalized. In fact whereas by construction
(\ref{Pif}) is normalized (provided the phase space density is normalized),
(\ref{PiJ}) is not automatically normalized for an arbitrary $\chi$, and can
actually be non-normalizable as in this example.
Positive, covariant, and normalized operational TOA distributions
will be obtained in the following section.

\section{Arthurs and Kelly model for
time of arrival and energy}
In the previous section TOA distributions have been defined indirectly
{}from position-momentum quasi-distributions.  Similar energy
distributions can also be obtained indirectly, as pointed out in
\cite{KW99}.  However, an Arthurs and Kelly type of measurement can be
used to find operationally TOA and energy  distributions in a
{\em direct} way: instead of the
conjugate variables position and momentum, we will describe the
measurement process in terms of the conjugate variables energy $E$ and
time of arrival $T$.  (In this section we shall use the somewhat
unusual notation $\wh E$ for the free motion Hamiltonian of the
particle $\wh E=\wh{H}=\wh{p}^ 2/(2m)$ to keep the notation in
parallel with the $\{x,p\}$ measurement model.)  In our $\{E,T\}$ model the
operator $\wh T$ is taken as the time operator introduced by Aharonov
and Bohm \cite{AB},
\beq\label{opT}
\wh T=-\frac{m}{2}\left(\wh{x}\frac{1}{\wh p}+\frac{1}{\wh p}\wh{x}\right)
\,,
\eeq
by symmetrizing the classical expression, $-m x/p$, for the time of
arrival at $X=0$, computed from $t$, 
of a particle that at time $t$ has position $x$ and
momentum $p$.  This operator is not self-adjoint but maximally
symmetric, see a detailed discussion in \cite{EM}.  Its (generalized)
eigenfunctions in momentum representation are given by (\ref{p|T}).
The $\{E,T\}$ Arthurs and Kelly measurement process is based on pointer
and particle operators, $\wh{\mu}_T, \wh\mu_E, \wh\pi_T, \wh\pi_E,
\wh{E}$, and $\wh T$, parallel to the set used for $x$ and $p$, and
related by similar commutation relations,
\beq\label{ih}
[\wh E,\wh T]=[\wh {\mu}_E,\wh {\pi}_E]=[\wh {\mu}_T,\wh {\pi}_T]=i\hbar\,,
\eeq
with all other commutators being zero.

The energy operator is  bounded from below.
As explained later it is also necessary to require that the operator
$\wh{\pi}_T$ be bounded from below.
Another difference with the $\{x,p\}$ case is the degeneracy,
associated  with positive and negative momenta, of
the spectra of $\wh E$ and $\wh T$. We shall keep the pointer variables
non-degenerate for simplicity although a more detailed model including
pointers sensitive to the degeneracy index is also possible.

The evolution operator describing the measurement process is
now given by
\beq
\wh{U}_{ET}=e^{-\frac{i}{\hbar}(\wh{\pi}_T \wh{T}+\wh{\pi}_E \wh{E})}\,.
\eeq
Our purpose is yet again to obtain the probability distribution for the
result of the measurement,
\beq\label{rhoet}
\rho(\mu_E, \mu_T)=\sum_\alpha \into dE\, |\la E, \alpha,
{\mu}_E, {\mu}_T|\wh{U}_{ET}|
\psi\otimes\psi_{ap}\ra|^2\,.
\eeq
In Sec. II, in the $\{x,p\}$ phase space,
the error
operators and all their relations were obtained using the
definitions of final and initial operators, the commutation relations
(\ref{ihxp}), and expression (\ref{rela}). In the $\{E,T\}$
phase space similar relations are also valid, and
(\ref{ret}) to (\ref{fin})
hold  by substituting $E$ for $x$ and $T$ for $p$.

In order to construct a meaningful Arthurs and Kelly model for TOA and
energy, it is useful to analyze first the simple, but not trivial,  von Neumann
measurement model of $\wh{T}$, corresponding to the operator
$\exp{-i\wh{\pi}_T \wh{T}/\hbar}$.  Note that the basis
$\{|T,\alpha\ra\}$ is complete but non-orthogonal, because
$\wh{T}$ is not self-adjoint.  However, the intermediate
computations are mostly carried out in the $\{|E,\alpha\rangle\}$
basis, which is indeed complete and orthogonal, since it corresponds
to the spectral decomposition of a self-adjoint operator.  Using the
overlap
\beq\label{et}
\la E,\alpha|T,\alpha'\ra
=\frac{1}{h^{1/2}}e^{iET/\hbar}\delta_{\alpha\alpha'},
\eeq
we find
\beq
\la E,\alpha,\pi_T|e^{-\frac{i}{\hbar}\wh{\pi}_T\wh{T}}  E',
\alpha',\pi_T'\ra=
\delta(E-\pi_T-E')\delta_{\alpha\alpha'} \delta(\pi_T-\pi_T')\,.
\eeq
When integrating over $E$, which is a positive variable, the energy delta
function cannot be satisfied unless $E'+\pi_T$ is positive.
This means that the exponential displaces the energy of the energy
eigenstate as long as $E'+\pi_T$ remains positive, but annihilates
the state otherwise,

\beq\label{vN}
e^{-\frac{i}{\hbar} {\pi}_T\wh{T}}| E',\alpha'\ra=
\Theta(E'+\pi_T)|E'+\pi_T,\alpha'\ra\,.
\eeq
It would be more rigurous to deal with wave packets in the energy
representation, given that $\hat T$ is not self-adjoint.  However, the
results we will be using are actually unchanged, so there is no need
to insist on this fact.  In order to guarantee that the exponential
operator is isometric (i.e.  preserves norm), we shall restrict the
spectrum of $\pi_T$ to the positive half line.  $\wh{\pi}_T$ and
$\wh{\mu}_T$ are thus a conjugate pair of operators similar to
$\wh{E}$ and $\wh{T}$ except for the degeneracy.  In particular, this
means that $\wh{\mu}_T$ is not self-adjoint, because of Pauli's
theorem \cite{Pauli,EM}.
Using similar arguments for each of the exponentials in
\beq
\wh{U}_{ET}=e^{-\frac{i}{\hbar}(\wh{\pi}_T \wh{T}+\wh{\pi}_E \wh{E})}
=e^{-\frac{i}{2\hbar}\wh{\pi}_T\wh{\pi}_E}
e^{-\frac{i}{\hbar}\wh{\pi}_T \wh{T}}e^{\frac{i}{\hbar}\wh{\pi}_E
\wh{E}}\,,
\eeq
it is easily proved that $\wh{U}_{ET}$ is isometric provided that
$\wh{\pi}_T$ is positive, and irrespective of the (real) spectrum
chosen for $\wh{\mu}_E$.  Indeed, there would be two different possible
models, depending on whether $0\le \mu_E< \infty$ or
$-\infty<\mu_E<\infty$ is chosen.  The bounded case has the advantage
of strictly positive values, mimicking the restriction on values of
$E$, but the disadvantage that initial apparatus states near the
origin cannot be Gaussian nor symmetrically peaked around $\mu_E=0$.
For the unbounded case, the initial state may be located symmetrically
around the origin, however, the price of there being negative values has to
be
paid.  We shall work out the second possibility hereafter, but, with
little changes, the analysis of the alternative apparatus may be also
carried out.  In particular, the important property of time covariance is
valid for both cases.  Let us first see how $\wh{U}_{ET}$ acts on a
convenient basis for the composite system:
\beq
\wh{U}_{ET}|E,\alpha,\mu_E,\pi_T\ra
=\bigg|E+\pi_T,\alpha,\mu_E+E+\frac{\pi_T}{2},
\pi_T\bigg>\,.
\eeq
By inserting a resolution of the identity in this basis one finds
\beq
\la E,\alpha,\mu_E,\pi_T|\wh{U}_{ET}|\psi\otimes\psi_{ap}\ra=
\bigg< E-\pi_T,\alpha,\mu_E-E+\frac{\pi_T}{2},
\pi_T\bigg|\psi\otimes\psi_{ap}\bigg>\Theta(E-\pi_T)\,,
\eeq
whose Fourier transform (to change from
$\pi_T$ to the conjugate variable $\mu_E$) is, using now the basis of
retrodictive and predictive energy errors for the apparatus state,
and $E'=E-\pi_T$,
\beq
\la E,\alpha,\mu_E,\mu_T|\wh{U}_{ET}|\psi\otimes\psi_{ap}\ra=
\frac{1}{h^{1/2}}\int_0^E dE' e^{\frac{i}{\hbar}[\mu_T(E-E')]}
\la\mu_E-E',\mu_E-E|\psi_{ap}\ra \la E',\alpha|\psi\ra\,.
\eeq
The upper limit reflects the fact that the measurement always shifts the
particle's energy ``upwards''.
Any resulting energy component $\la E|$ can only have
contributions from the lower energy states.
{}From here the operational joint time-of-arrival and energy
distribution takes the form
\beqa
\rho[\mu_E,\mu_T;\psi(t=0)]&=&\frac{1}{h}\sum_\alpha\into
dE\int_0^E dE' \int_0^E dE''
e^{\frac{i}{\hbar}\mu_T(E''-E')}
\nonumber\\
&\times& \la \mu_E-E',\mu_E-E|\psi_{ap}\ra\la \psi_{ap}|\mu_E-E'',\mu_E-E\ra
\la E',\alpha|\psi(0)\ra\la\psi(0)|E'',\alpha\ra\,.
\label{ETdist}
\eeqa
Let us now examine some properties of this distribution:
It is time covariant as can be easily seen from
(\ref{ETdist}):
\beqa
&&\rho[\mu_E,\mu_T-t;\psi(t)]=
\sum_\alpha \frac{1}{h} \into
dE\int_0^E dE' \int_0^E dE''
e^{\frac{i}{\hbar}(\mu_T-t)(E''-E')}
\nonumber\\
&\times&
\la \mu_E-E',\mu_E-E|\psi_{ap}\ra\la \psi_{ap}|\mu_E-E'',\mu_E-E\ra
\la E',\alpha|e^{-iE't/\hbar}|\psi(0)\ra\la\psi(0)| e^{iE''t/\hbar}
|E'',\alpha\ra
\nonumber\\
&=&\rho[\mu_E,\mu_T;\psi(0)]\,,
\eeqa
because of the cancellation of the $t$-dependent exponentials.
It is also possible to write it in a form similar, but not identical,
to Eq. (\ref{rhoxp}).
To this end we shall  define the following apparatus dependent
object in the basis of retrodictive error of energy,
\beq\label{cosa}
\la a|\wh{\varrho}_{ap}(c)|b\ra\equiv\into dE\, \la a, c-E|\psi_{ap}\ra
\la \psi_{ap}|b,c-E\ra\Theta(E-c+a)\Theta(E-c+b)\,.
\eeq
Changing variables to the half-sum and difference,
\beqa\label{cv}
s&=&\frac{E'+E''}{2}
\nonumber\\
y&=&E''-E'\,,
\eeqa
we can write (\ref{ETdist}) as
\beqa
\rho(\mu_E,\mu_T)&=&\sum_\alpha\frac{1}{h}\into ds\intf dy\,
e^{i\mu_T y/\hbar}\la \mu_E-s+y/2|\wh{\varrho}_{ap}(\mu_E)|\mu_{E}-s-y/2\ra
\nonumber\\
&\times&
\la s-y/2,\alpha|\psi\ra\la\psi|s+y/2,\alpha\ra\,\Theta(s-y/2)\Theta(s+y/2)\,.
\label{lll}
\eeqa
Alternatively,
\beq\label{rhoww}
\rho(\mu_E,\mu_T)=\sum_\alpha\into dE\intf dT\,
W_{\mu_E}(\mu_E-E,\mu_T-T)w_{\alpha,\alpha}(E,T)\,,
\eeq
where
\beq\label{wsy}
w_{\alpha,\alpha'}(E,T)=\frac{1}{h}\intf dy\,
e^{iTy/\hbar}\la E-y/2,\alpha|\psi\ra\la\psi|E+y/2,\alpha'\ra
\Theta(E-y/2)\Theta(E+y/2)\,,
\eeq
and
\beq\label{wap}
W_{\mu_E}(\mu_E-E,\mu_T-T)=
\frac{1}{h}\intf dy\, e^{-i(\mu_T-T)y/\hbar}
\la \mu_E-E-y/2|\wh{\varrho}_{ap}(\mu_E)|\mu_E-E+y/2\ra\,,
\eeq
as can be checked by substitution of (\ref{wsy}) and (\ref{wap}) into
(\ref{rhoww}), and renaming  $s=E$ in (\ref{lll}).
$W_{\mu_E}$ is an apparatus dependent window function
for the energy-TOA distribution of the particle.  Even though
(\ref{wap}) and (\ref{cosa}) provide an explicit expression, its
interpretation is not as simple as the corresponding phase space
window function for the $\{x,p\}$ case of Sec.  II. Because of the
lower energy bound of the energy, $\wh{\varrho}_{ap}(\mu_E)$ is not,
in general, the reduced density operator of the apparatus for the
retrodictive errors.  The filtered function, $w_{\alpha,\alpha}(E,T)$,
has a simpler content as an $\alpha,\alpha$ diagonal component of the
energy-TOA Wigner matrix for the particle, Eq.  (\ref{wsy}).
Remarkably, no interference term with $\alpha\ne\alpha'$ contributes
to (\ref{rhoww}), a feature shared with Kijowski's time-of-arrival
distribution.  As a matter of fact, by tracing over $\alpha$ and
integrating over $E$, the marginal of the particle's Wigner matrix is
nothing but Kijowski's distribution,
\beqa
\sum_\alpha \into dE\, w_{\alpha,\alpha}(E,T)&=&
\frac{1}{h}\sum_\alpha \into\! dE'\!\into\! dE'' e^{iT(E''-E')/\hbar}
\la E',\alpha|\psi\ra\la\psi|E'',\alpha\ra
\nonumber\\
&=&\sum_\alpha\la T,\alpha|\psi\ra
\la\psi|T,\alpha\ra=\Pi_K(T)\,,
\eeqa
as is readily seen by undoing  the change of variables displayed in
(\ref{cv}) and using Eq. (\ref{et}).
However, whereas in the von Neumann measurement the resulting distribution
$P(\mu_T)$ is
a smoothed version of Kijowski's distribution, see Appendix B,
the TOA marginal of the Arthurs and Kelly distribution obtained by
integrating (\ref{rhoww}) over $\mu_E$,  not only smooths  but also
distorts Kijowski's distribution. This is due to the
double dependence on $\mu_E$ of the apparatus dependent window function
$W_{\mu_E}(\mu_E-E, \mu_T-T)$: aside from
$w_{\alpha,\alpha}(E,T)$
the integral over $\mu_E$  leaves an extra $E$-dependent function.
This is one further peculiarity of the $\{E,T\}$ Arthurs and Kelly model
with respect to the $\{x,p\}$ model of Sec. II.
In the later, there is no double dependence
on $\mu_X$ in the apparatus window Wigner function, and the marginal
is simply a smoothed version of the quantum
mechanical distribution of the particle's position, without
additional distortion.

\section{Discussion}
In this work we have explored different ways to obtain time-of-arrival
(TOA) distributions ``operationally'', i.e., by means of models that
include the particle and additional degrees of freedom acting as
meters or sensors of the particle.  An indirect route is to {\it
define} the TOA distribution from a distribution of position and
momentum that is determined operationally.  The Arthurs and Kelly
measurement model is possibly the simplest model where such
distributions may be obtained.  We have shown however that in order to
obtain a time covariant result the step from the phase space
distribution to the TOA distribution is delicate, even crucial.
Furthermore, when we consider generic distributions of position and
momentum the essential property of time covariance proves to be rather
elusive.  In particular, we have proved that for the choice of
classical phase space function $\delta(T+mq/p)$ (that is, the choice
of Kocha\'nski and W\'odkiewicz, \cite{KW99}) there is no distribution
of position and momentum within a wide class for which the associated
TOA distribution is covariant in time.  Moreover, we have given yet
another instance of the well known fact that two classically
equivalent expressions may give different quantum distributions,
which, in the case at hand, are covariant or not.  Thus, we have
proposed a different classical phase space function, $\delta(q)|p/m|$,
for which covariance is always granted for a wide class of phase space
families of distributions parameterized by $T$, compare (\ref{Pif})
and (\ref{PiJ}).  Not only are the quantum distributions numerically
different from the classical ones, conceptually they are also worlds
apart: whereas the quantization of Kocha\'nski and W\'odkiewicz should
be more properly understood as the quantization, within the same
quantization scheme, of a family of classical functions, our proposal
is rather a family of quantization schemes of the same classical
function.

The question now to be posed is whether any of these different
indirect methods of obtaining TOA distributions can be selected as
being ``better'' in some suitable sense. From the theoretical point of
view, covariance and positivity seem to be minimal requirements,
therefore selecting positive flux ($\delta(q)|p/m|$) over trajectory
identification ($\delta(T+mq/p)$).

Nonetheless, in the case of the free particle, not even the
quantization of positive flux provides us with the ``best'' TOA
distribution, in that it may be not normalizable, as we have shown with an
example, and that for any quantization of positive flux the
variance of the distribution will be larger than for Kijowski's
distribution, which we have shown does not lie within a wide class of
quantizations of $\delta(q)|p/m|$.
We are in this manner led to propose a different operational
approach, in which time of arrival and energy are
directly the ``measured'' variables.  In this case time covariance
is automatically satisfied. Not only that: Kijowski's TOA distribution
can now be understood as a marginal distribution deriving from the Wigner
function, suitably generalized for the pair of variables energy and
time of arrival.
Operationally,
a simple TOA von Neumann measurement provides a smoothed Kijowski's
distribution whereas the marginal of the time-energy distribution
that results from the Arthurs and Kelly process is distorted
beyond a simple smoothing.
In both cases and for a  generic apparatus state
we see that the variance of the TOA distributions hence derived is
bigger than for
Kijowski's, as was only to be expected from the axiomatic derivation
of Kijowski's distribution, which selects $\Pi_K$ as the one with smallest
variance \cite{Kijowski}.

Additional to our central objective, namely, the study of operational
definitions of TOA distributions and their relation to ideal
measurements of time, we have also obtained a general result
concerning operational definitions of measurements of pairs of
conjugate variables when one of them has a bounded spectrum from
below.  For instance, $x$ and $p$ in the half line.
The na\"\i ve
construction of Wigner's function for such a case is actually valid,
even though $\wh{p}$ is not self-adjoint.  The reason for the validity of
the na\"\i ve expression is, however, far from simple.  Since $\wh{p}$ is
not
self-adjoint and admits no self-adjoint extension on the half-line
(technically, it is a maximally symmetric operator with $(1,0)$ defect
indices), there is no spectral decomposition available for it.  Even
so, we do have the next best thing, namely, a positive operator valued
measure (POVM) \cite{Busch,EM}, and this allows the construction of the
basis of operators required for the definition of Wigner's function.
Moreover, the probability density obtained as the marginal
distribution for the na\"\i vely defined Wigner function can be used
to reconstruct the whole POVM. To the best of our knowledge, the
applicability of the Wigner function formalism and its generalizations
has not been discussed heretofore in the literature for this case.

On a more general note, we would like to point out that even though
the world of ``operational models'' is in principle between the
ideal theories
(depending only on the particle's state) and actual experiments,
the connections are not always explicit or obvious.  We have in
particular emphasized that the phase space probability obtained
operationally by means of the position-momentum Arthurs and Kelly
process may be regarded as one particular (ideal) joint
quasi-distribution.  Many others exist, in particular the ones in the
family defined by Cohen, but their links with a particular operational
procedure are not generally as direct.  The fact that Kijowski's ideal
TOA distribution, the closest object to the classical TOA distribution
considering the common properties satisfied, is not obtained with any
of these phase space quasi-distributions indicates that there is a
limit to the flexibility of the class of quantization rules included
in Cohen's formalism, a warning flag worth considering when dealing
with other quantization problems.

The other connection, between operational results an actual experiments
is sometimes taken for granted, but in fact it is far from being
easily realized. The conjugate variables measured by
``Arthurs and Kelly experiments'' in quantum optics
are the two quadrature components
of the electric field strength of a single-mode radiation field,
not the position and momentum of a particle.
Nevertheless we expect a similar experiment with particles to be 
feasible soon (see \cite{PTW}).
In future work we also intend to relate the ideal TOA distributions with operational
models that describe a continuous (rather than instantaneous) measurement 
process, thus being closer to an actual time-of-flight experiment as emphasized 
by Aharonov et al. \cite{AOPRU}.

We thank Rick Leavens and D. Alonso
for useful discussions, and acknowledge support
by Gobierno Aut\'onomo de Canarias (PB/95), MEC (PB97-1482), CERION, and The
University of the Basque Country (project UPV 063.310-EB187/98)
\appendix
\section{Basic Cohen's phase space formalism}
Given an operator $\wh{G}$ and a density operator $\wh\varrho$ their
phase space representatives
\beqa\label{Frho}
F(x,p)&=&\frac{1}{4\pi^{2}}
\intf d\theta\intf d\tau \intf du\,
\bigg <u+\frac{\hbar\tau}{2}\bigg|\widehat{\varrho}
\bigg|u-\frac{\hbar\tau}{2}\bigg >
e^{-i[\theta(x-u)+\tau p]}\chi(\theta,\tau)
\\\label{gG}
g(q,p)&=&\frac{\hbar}{2\pi}\intf d\theta\intf d\tau\intf du\,
\bigg<u-\frac{\hbar\tau}{2}\bigg|\widehat{G}\bigg|u+
\frac{\hbar\tau}{2}\bigg>
\frac{e^{i[\theta (q-u)+\tau p]}}{\chi(\theta,\tau)}
\eeqa
are chosen such that
\beq\label{expec}
\la\widehat{G}(\widehat{x},\widehat{p})\ra=
{\rm tr} (\widehat{\varrho}\,\widehat{G})=
\intf dx\intf dp\, F(x,p)\, g(x,p).
\eeq
Conversely, given two space space representatives for the state and
dynamical variable, their operators are obtained by
\beqa\label{Gg}
\label{rhoF}
\widehat{\varrho}=\frac{\hbar}{2\pi}
\intf dx\intf dp\intf d\theta\intf d\tau\,
F(x,p)\chi(\theta,\tau)^{-1}e^{i[\theta(x-\widehat{x})+
\tau (p-\widehat{p})]}
\\
\widehat{G}(\widehat{x},\widehat{p})=
\frac{1}{4\pi^{2}}\intf dx\intf dp\intf d\theta\intf d\tau\,
g(x,p)e^{-i(\theta x+\tau p)}\chi(\theta,\tau)
e^{i(\theta\widehat{x}+\tau\widehat{p})}.
\eeqa

\section{von Neumann measurement of $\wh{T}$}
Using (\ref{vN})  the amplitude that results from
the von Neumann  measurement of the time operator $\wh T$
is given, for a semibounded $\wh{\pi}_T$, by
\beq
\la T,\alpha,\mu_T|e^{-i\wh{\pi}_T\wh{T}/\hbar} \psi(t)\otimes\psi_{ap}\ra
=
\la \mu_T-T|\psi_{ap}\ra \la T,\alpha|\psi(t)\ra\,.
\eeq
Since there is only one pointer  the wave function of the apparatus,
$\psi_{ap}$, depends only
on one variable here. The probability to find $\mu_T$ in a
 measurement carried out at $t$ is a smoothed version
of Kijowski's distribution, see (\ref{Kijo}),
\beq
P(\mu_T;t)=\intf dT\, |\la\mu_T-T|\psi_{ap}\ra|^2\, \Pi_K(T;t)\,,
\eeq
which is covariant $[P(\mu_T-t';t+t')=P(\mu_T;t)]$
because of the covariance of $\Pi_K$.

%
%

%
\newpage
FIGURE CAPTIONS:\\

{\bf Figure 1}: $\Pi_{KW}(T-t;t)$ versus $T$:
$t=-0.2$ (long dashed line); $t=-0.1$ (short dashed line).
$\Pi_{\tilde{J}}(T-t;t)$ versus $T$ for any $t$, (solid line).
In all cases  $x_0=-2.5$,
$k_0=10$, $\sigma=\delta=0.1$, and $\hbar=m=1$ (atomic units).
The bumps on the left/right are essentially due to contributions from 
negative/positive momenta.


\begin{references}
%
\bibitem{AK} E. Arthurs and J. L. Kelly, Bell Syst. Tech. J. {\bf 44},
1153 (1965).
%
\bibitem{Leonh} Leonhardt U. {\it Measuring the Quantum State of
Light}, (Cambridge University Press, Cambridge, 1997).
%
\bibitem{Braun} S. L. Braunstein, C. M. Caves and G. J. Milburn, Phys.
Rev A {\bf 43}, 1153 (1991).
%
\bibitem{Sten1} S. Stenholm, Ann. Phys. {\bf 218}, 233 (1992).
%
\bibitem{Wod2} P. Kocha\'nski and K. W\'odkiewicz, Rep Math. Phys. {\bf
40}, 245 (1997).
%
\bibitem{App1} D. M. Appleby, Int. J. Theor. Phys. {\bf 37}, 1491
(1998).
%
\bibitem{App2} D. M. Appleby, J. Phys. A {\bf 31}, 6419 (1998).
%
\bibitem{Muy1} H. Martens and W. M. de Muynck, Found. Phys. {\bf 20},
357 (1990).

\bibitem{Muy2} W. M. de Muynck, quant-ph 9901010.

\bibitem{Busch85} P. Busch, Int. J. Theor. Phys. {\bf 24}, 63 (1982).

\bibitem{AAU} Y. Aharonov, D. Z. Albert, and C. K. Au, Phys. Rev. Lett.
{\bf 47}, 1029 (1981).
%
\bibitem{OR} R. F. O'Connell and A. K. Rajagopal, Phys. Rev. Lett. {\bf
48}, 525 (1982).
%
\bibitem{Wod} K. W\'odkiewicz, Phys. Rev. Lett. {\bf 52}, 1064 (1984).
%
\bibitem{Hall} J. J. Halliwell, Physical Review D {\bf 46}, 1610 (1992).
%
\bibitem{Sten2} P. T\"{o}rm\"{a}, S. Stenholm and I. Jex, Phys. Rev. A
{\bf 52}, 4812 (1995).
%
\bibitem{Agar} G. S. Agarwal and S. Chaturvedi, Phys. Rev. A {\bf 49},
R665 (1994).
%
\bibitem{VN} J. von Neumann, {\it Mathematical Foundations of Quantum
Mechanics} (Princeton
University Press, Princeton, 1955).
%

\bibitem{Davies} E. B. Davies, {\it Quantum Theory of Open Systems}
(Academic Press, New York, 1976).
%
\bibitem{KW99} P. Kocha\'nski and K. W\'odkiewicz,
Phys. Rev. A {\bf 60}, 2689 (1999).
%
\bibitem{Mugarev} J. G. Muga, R. Sala, and J. P. Palao, Superlattices
and Microstructures {\bf 23}, 833 (1998). See also J. G. Muga and C. R.
Leavens, {\it Arrival time in quantum mechanics,} in preparation.
%
\bibitem{MLP} J. G. Muga, C. R. Leavens and J. P. Palao, Phys. Rev. {\bf 58},
4336 (1998)
%

\bibitem{Finkelstein} J. Finkelstein, Phys. Rev. A {\bf 59}, 3218 (1999).

\bibitem{Toller} M. Toller, Phys. Rev. A {\bf 59}, 960 (1999).

\bibitem{Kijo99} J. Kijowski, Phys. Rev. A {\bf 59}, 897 (1999).

\bibitem{ORU} J. Oppenheim, B. Reznik, and W. G. Unruh, Phys. Rev. A
{\bf 59}, 1804 (1999).

\bibitem{Delgado} V. Delgado, Phys. Rev. A {\bf 59}, 1010 (1999).

\bibitem{MPL} J. G. Muga, J. P. Palao, C. R. Leavens, Phys. Lett. A
{\bf 253}, 21 (1999).

\bibitem{EM} I. Egusquiza and J. G. Muga, Phys, Rev. A (1999), accepted,
quant-ph 9904055.

\bibitem{BEMS} A. D. Baute, R. Sala Mayato, J. P. Palao, J. G. Muga and
I. L. Egusquiza,
Phys. Rev. A {\bf 61} (2000), accepted, quant-ph 9904055.

\bibitem{LJPU} J. Le\'on, J. Julve, P. Pitanga and F. J. Urr\'\i es,
quant-ph 9903060.

\bibitem{Galapon} E. Galap\'on, quant-ph 9908033.

\bibitem{Kijowski} J. Kijowski, Rep. Math. Phys. {\bf 6}, 362 (1974).
%
\bibitem{Bohm} C. R. Leavens, Phys. Rev. A {\bf 58}, 840 (1998).
%
\bibitem{AOPRU} Y. Aharonov, J. Oppenheim, S. Popescu, B. Reznik,
and W. G. Unruh, Phys. Rev. A {\bf 57}, 4130 (1998).
%
\bibitem{Halliwell} J. Halliwell, quant-ph/9805057, to appear in
Prog. Th. Phys. 102 (1999).
%
%
\bibitem{MBM} J. G. Muga, S. Brouard and D. Mac\'\i as, Annals of Physics
{\bf 240}, 351 (1995).
%
\bibitem{Busch} P. Busch, M. Grabowski and P. J. Lathi, {\it
Operational Quantum Mechanics} (Springer-Verlag, Berlin, 1995).
%
\bibitem{Cohen1} L. Cohen, Jour. Math. Phys. {\bf 7}, 781 (1966).
%
\bibitem{Cohen2} L. Cohen, {\it Time-Frecuency Analysis} (Prentice
Hall, New York, 1995).
%
\bibitem{Dav} E. B. Davies, {\it Quantum Theory of Open System} (Academic
Press, New York, 1976).
%
%
\bibitem{Werner} R. Werner, J. Math. Phys. {\bf 27}, 793 (1986).
%
%
\bibitem{Allc} G. R. Allcock, Ann. Phys. (NY) {\bf 53}, 253 (1969); G.
R. Allcock, Ann. Phys. (NY) {\bf 53}, 286 (1969); G. R. Allcock, Ann.
Phys. (NY) {\bf 53}, 311 (1969).
%
\bibitem{Brac} A. J. Bracken and G. F. Melloy, J. Phys. {\bf A27},
2197 (1994).
%
\bibitem{class} We only consider $\chi$ kernels which are state-independent 
so that the associated quantization rules are also state independent, and the 
phase space representatives of the quantum state depend bilinearly on the wave function. 
Other possibilities have 
been considered by Cohen and coworkers, see  L. Cohen and Y. Zaparovanny,
J. Math. Phys. {\bf 21}, 794 (1980).
%
\bibitem{AB} Y. Aharonov and D. Bohm, Phys. Rev. A {\bf 122}, 1649
(1961).
%
%
\bibitem{Pauli} W. Pauli, Die allgemeinen prinzipien der Wellenmechanik.
Handbuch
der Physik, ed. S. Fl\"uge, vol. V/1, Springer-Verlag (1958); English
version: General
Principles of Quantum Mechanics, Springer-Verlag, Berlin (1980) (p.
63).
%
\bibitem{PTW} W. L. Power, S. M. Tan, and M. Wilkens,
J. Mod. Opt. 44 (1997) 2591.
%

\end{references}
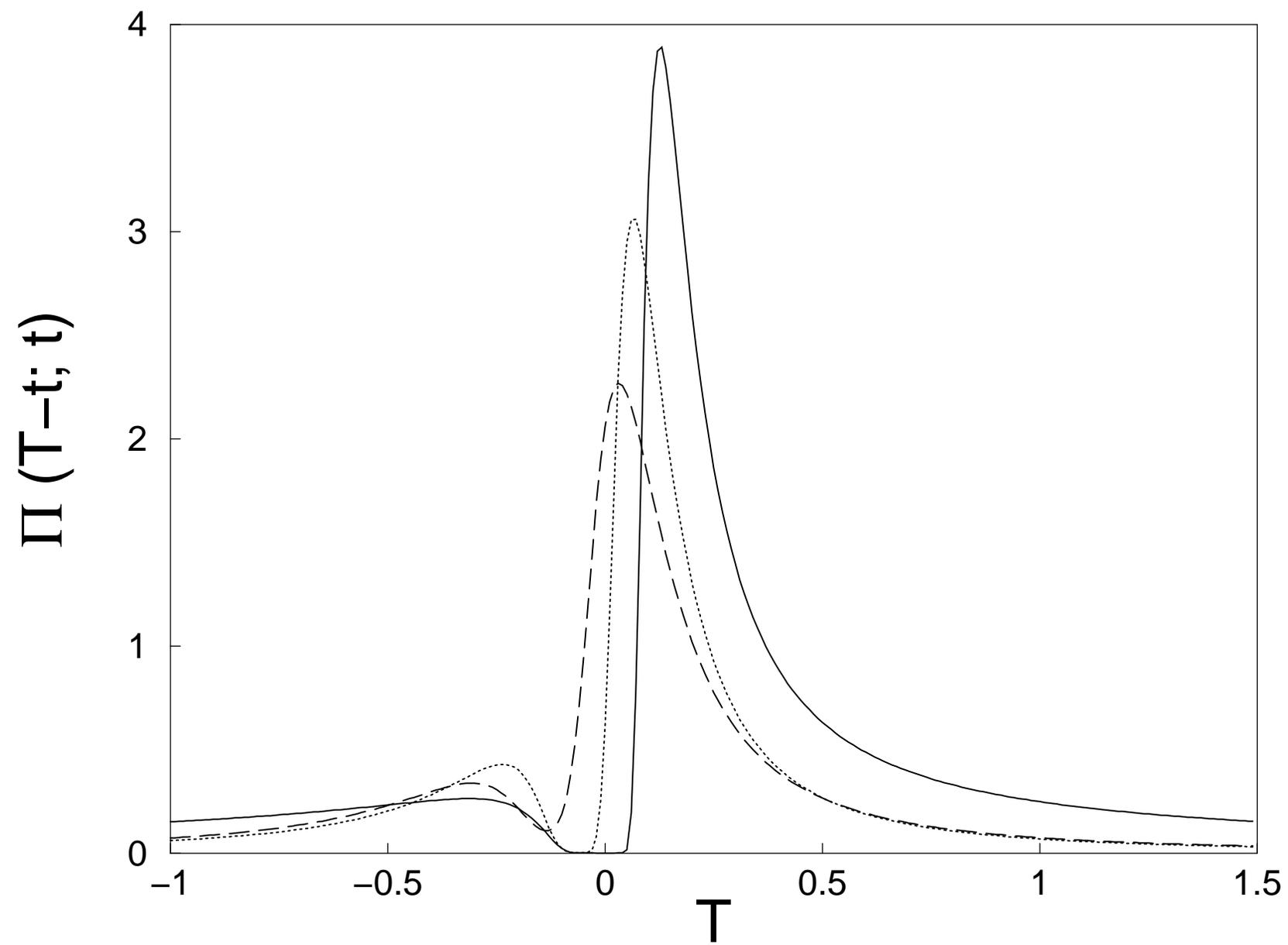
\end{document}